\documentclass[aps,prl,twocolumn,superscriptaddress,preprintnumbers]{revtex4-1}
\usepackage{amsmath}
\usepackage{amssymb}
\usepackage[section]{placeins}
\usepackage{units}
\usepackage{graphicx}

\newcommand{\pt}{p_\perp}
\newcommand{\kt}{k_\perp}

\newcommand{\alphas}{\alpha_\text{s}}
\newcommand{\ti}{T_\text{i}}
\newcommand{\taui}{\tau_\text{i}}
\newcommand{\tc}{T_\text{c}}
\renewcommand{\d}{\text{d}}
\newcommand{\jewel}{\textsc{Jewel}\ }

\begin{document}

\preprint{IPPP/11/41}
\preprint{DCPT/11/82}
\preprint{MCnet-11-26}
\preprint{CERN-PH-TH/2011-299}

\title{Explaining jet quenching with perturbative QCD alone}

\author{Korinna~C.~Zapp}
\email{k.c.zapp@durham.ac.uk}

\author{Frank Krauss}
\affiliation{Institute for Particle Physics Phenomenology, Durham University, Durham\ DH1\,3LE, UK}

\author{Urs~A.~Wiedemann}
\affiliation{Department of Physics, CERN, Theory Unit, CH-1211 Geneva 23}

\date{\today}

\begin{abstract}

We present a new formulation of jet quenching in perturbative QCD beyond the eikonal 
approximation.  Multiple scattering in the medium is modelled through infra-red-continued 
($2 \to 2$) scattering matrix elements in QCD and the parton shower describing further 
emissions. The interplay between these processes is arranged in terms of a formation time
constraint such that coherent emissions can be treated consistently. 
Emerging partons are hadronised by the Lund string model.
Based on this picture we obtain a good 
description of the nuclear modification factor at \textsc{Rhic} and 
\textsc{Lhc} and predict its behaviour at very large $\pt$.
\end{abstract}

\maketitle

High-momentum transfer processes, such as single inclusive hadron spectra, jet-like 
particle correlations and reconstructed jets, have been advocated for a long time as precision 
tools for studying the dense QCD matter produced in ultra-relativistic nucleus-nucleus 
collisions. The rates and distributions of such processes can be benchmarked in the 
absence of medium effects both experimentally (e.g.\ through their measurements in 
proton-proton collisions) and theoretically (due to their perturbative calculability). 
In nucleus-nucleus collisions at \textsc{Rhic} and at the \textsc{Lhc}, they show generic and strong 
medium modifications commonly referred to as {\it jet quenching}. This motivates 
using such hard processes as well-calibrated {\it hard probes} of dense QCD matter. 
A central challenge for theoretical work is to relate, in a quantitatively controlled 
and unambiguous way, the observed medium modifications to fundamental properties of 
the QCD matter that induces them. In this context, applying a perturbative description 
plays a central role.

The first measurements of jet quenching signatures at the \textsc{Lhc} sparked an intense debate
on the extent to which a purely perturbative description of jet quenching is feasible.
On the one hand, it seems plausible that several features of partonic in-medium propagation
may resist a perturbative formulation. This is the case, for instance, for effects of
colour flow between jet and medium that modify the non-perturbative late-time dynamics of
jet hadronisation. On the other hand, several perturbatively motivated jet quenching
models~\cite{Horowitz:2011gd,Renk:2011gj,Chen:2011vt,Majumder:2011uk}
resting on the eikonal limit~\cite{Baier:1996sk,Zakharov:1997uu,Wiedemann:2000za,Wang:2001ifa,Gyulassy:2000er,Arnold:2002ja}
have been shown to reproduce central features of the nuclear modification factors at \textsc{Rhic} and at the \textsc{Lhc}.
However, the eikonal limit is known to be plagued by large numerical uncertainties~\cite{Armesto:2011ht},
and this complicates an assessment of whether a purely perturbative description can account for jet quenching.

Here, we formulate a conceptually new framework based on our understanding 
of perturbative QCD in non-eikonal kinematics and to document its first implementation in the
medium-modified final state parton shower \textsc{Jewel}. This formulation treats the scattering 
of an energetic parton off a medium constituent as a standard $2\to 2$ partonic scattering process 
described by QCD matrix elements and valid in the full perturbative
regime. Further following the standard description of all hard scattering events at colliders, large logarithms associated
to collinear singularities in real emission matrix elements are resummed by the parton shower. This
provides a systematically improvable approximation to extra gluon emissions that also includes radiation off the
scattering centre and generates naturally configurations with any number of additional patrons.
The parton shower thus introduces radiative energy loss, while the matrix elements
model the collisional one.  The ambiguity between elastic and
inelastic scattering is thus resolved, and the distinction between vacuum and medium induced radiation becomes
obsolete. Treating all emissions fundamentally in the same way introduces in a natural way the
 interplay of radiation from different sources.

In the following we discuss how this idea can be implemented into a Monte Carlo event generator 
allowing for a democratic treatment of all partonic fragments, and, at the same time, accounting 
in a local and probabilistic formulation not only for exact four momentum conservation including
all recoils, but also with known accuracy~\cite{Zapp:2008af,Zapp:2011ya} for the dominant quantum 
interference effect in medium-induced gluon radiation, the so-called non-Abelian 
Landau-Pomerantchuk-Migdal (\textsc{Lpm}) effect. 

\medskip

The parton shower evolution is governed by the Sudakov form factor, which can be interpreted 
as the probability for having no emission between two scales $Q^2$ and $Q_0^2$.
\begin{equation}
\label{eq:sudakov}
\mathcal{S}_a (Q^2,Q_0^2)
   =  \exp\left[ - \int \limits_{Q_0^2}^{Q^2}\!\! \frac{\d q^2}{q^2} \int\!\! \d z 
   \frac{\alphas (\kt^2)}{2\pi} {\sum_b}  {\hat P}_{ba}(z) \right]
\end{equation}
The functions $\hat P_{ba}(z)$ are the (unregularised) Altarelli-Parisi splitting functions 
describing the energy fraction $z$ taken by parton $b$ in the splitting $a\to b+c$. The coupling 
is evaluated at the transverse momentum of the splitting; it is given by $\kt^2 \simeq (1-z)q^2$ 
for initial state and by $\kt^2 \simeq z(1-z)q^2$ for final state emissions.

In order to include secondary scatters and initial state radiation off such collisions, we 
introduce `partonic pdf's' $f_j^i(x,Q^2)$, which analogously to the proton pdf's describe -- 
at leading order -- the probability that a parton $j$ with energy fraction $x$ has been radiated 
from a parton $i$ when the latter is resolved at a scale $Q^2$. These pdf's are also employed 
in the typical backward evolution of initial state radiation, as known, e.g., from simulating 
proton-proton collisions. Since the centre-of-mass energies of scatterings in the medium are 
typically rather small, it is unlikely that more than one initial state splitting is initiated 
in secondary scatters. In this approximation\footnote{This approximation can easily be improved numerically.} 
the partonic pdf's can be integrated analytically, for instance
\begin{align}
&f_{\text{q}}^{\text{q}}(x,Q^2)
 =  \mathcal{S}_{\text{q}}(Q^2,Q_0^2) \delta(1-x) \nonumber  \\
   & \qquad + \int\limits_{Q_0^2}^{Q^2}
\! \frac{\d q^2}{q^2}\, \mathcal{S}_{\text{q}}(Q^2,q^2)
\, \frac{\alphas((1-x)q^2)}{2\pi} {\hat P}_{\text{qq}}(x)\, ,
\end{align}
and analogously for the other partonic pdf's.

The interaction of the parton shower with the parton constituents in the medium depends on 
properties of the medium. We specify the cross sections for $2 \to 2$ processes as
\begin{equation}
\label{eq:crosssection}
\sigma_i(E,T) = 
\hspace{-0.25cm}  \int\limits_0^{|\hat t|_{\text{max}}(E,T)}\!\!\!\!\! \d |\hat t|\!\! 
 \int\limits_{x_{\text{min}}(|\hat t|)}^{x_{\text{max}}(|\hat t|)} \!\!\!\!\! \d x
\sum_{j \in \{\text{q,\=q,g}\}} \!
f_j^i(x, \hat t) \frac{\d \hat \sigma_j}{\d \hat t}(x\hat s,|\hat t|) \,,
\end{equation}
where the pdf takes into account possible initial state radiation off the energetic projectile.
Note that here, implicitly, we neglected a similar evolution experienced by the target, i.\,e.\ 
there is no initial state radiation off the medium parton.
The maximum momentum transfer $|\hat t|_\text{max}$ is determined by the kinematics
of the scattering. Neglecting the scattering centre's momentum one finds 
$|\hat t|_\text{max} = 2 m_\text{s}(T) [E_\text{p} - m_\text{p}]$, where $m_\text{s}(T)$
stands for the (temperature dependent) scattering centre's mass and $E_\text{p}$ and $m_\text{p}$ are
the projectile parton's energy and mass, respectively. The boundaries on the 
$x$-integral are obtained from the requirement that $\kt^2 \ge Q_0^2/4$ and are
given by $x_{\text{min}}(|\hat t|) = Q_0/(4|\hat t|)$ and 
$x_{\text{max}}(|\hat t|) = 1 - Q_0/(4|\hat t|)$.
For the partonic cross section we keep leading terms only, but we regularise 
them with a Debye mass $\mu_\text{D} \approx 3 T$. They therefore read
\begin{align}
 \frac{\d \hat \sigma}{\d \hat t}(\hat s,|\hat t|) & = C_{\text{R}}
\frac{\pi}{\hat s^2} \alphas^2(|\hat t| + \mu_{\text{D}}^2)\frac{\hat s^2 + (\hat
s-|\hat t|)^2}{(|\hat t| + \mu_{\text{D}}^2)^2} \nonumber \\
& \longrightarrow C_{\text{R}}
2 \pi \alphas^2(|\hat t| + \mu_{\text{D}}^2)\frac{1}{(|\hat t| +
\mu_{\text{D}}^2)^2} \,.
\label{cross}
\end{align}

Each parton emitted in subsequent parton showering has a finite formation time that is 
parametrically given by $\tau \approx \omega/\kt^2 \approx (x)E/Q^2$.
When the formation time is larger than the mean free path in the medium,
emissions from all scattering centres within in the formation time interfere
destructively giving rise to the non-Abelian Landau-Pomerantchuk-Migdal (\textsc{Lpm})
effect. This has been incorporated following the algorithm of
\cite{Zapp:2008af,Zapp:2011ya}, with the only difference that in the current
framework gluon formation times cannot overlap. Compared to the original algorithm, 
this amounts to a mild and well understood suppression of induced radiation 
compared to the analytic result without changing the shapes of the gluon 
distributions~\cite{beyondbdmps}.

When emissions from different sources, i.e.\ initiated by different hard
scatterings (where one may be the 'vacuum' parton shower) compete with each other, 
it is solely the gluon with the shortest formation time that will be emitted.
This criterion is enough to regulate the interplay of all (vacuum and medium
induced) emissions. 

\medskip

To obtain the example results below, the following set-up has been used:
In all cases, the final state parton shower is generated by \jewel~\cite{Zapp:2008gi},
where elastic scattering and medium-induced radiation is implemented as discussed above. 
Initial dijet production is simulated using the matrix elements and initial state parton 
shower of \textsc{Pythia}~6.4~\cite{Sjostrand:2006za}, the latter running in a
virtuality-ordered mode to provide the best fit to the \jewel simulation. In all results, 
the strong coupling is running at one loop with $\alphas(m_{\text{Z}})=0.128$, consistent with findings of other
leading 
order calculations. The infra-red cut-off of the parton shower was chosen as $Q_0 = \unit[1.5]{GeV}$
to yield a good description of \textsc{Lep} data. 
The \textsc{Cteq6l1}~\cite{Pumplin:2002vw} pdf's as provided by 
\textsc{Lhapdf}~\cite{Whalley:2005nh} are used; simulating nuclear collisions the nuclear 
modification of \textsc{Eps09lo}~\cite{Eskola:2009uj} is employed. The medium is modelled 
using a simple variant of the Bjorken model ~\cite{Zapp:2005kt}, which roughly accounts 
for the main features of the produced matter. It describes the boost-invariant 
longitudinal expansion of an ideal quark-gluon-gas. The transverse profile is chosen such that the 
energy density is proportional to the density of wounded nucleons. 
\textsc{Jewel} can in principle be interfaced with any medium model.
For the hadronisation the strings are built 
by \jewel and then handed over to the \textsc{Pythia} hadronisation 
routine~\cite{Andersson:1983ia,Sjostrand:1984ic}. The colour flow and consequently the
strings can be constructed in different ways. The default choice reported here is to treat 
the struck scattering centres as if they were gluon emissions. The scattering centre is then 
always colour-connected to the projectile, before the parton shower sets in. 

In principle, the modelling of interactions in the medium is free of parameters, as 
the parton shower and its infra-red cut-off $Q_0$, hadronisation and the couplings are fixed from $e^+e^-$ 
and $pp$ collisions. In practice, however, the infra-red regulator in the cross section (\ref{cross}) 
adds an additional freedom. Here, we choose $\mu_{\text{D}} \approx 3 T$, in accordance with 
parametric estimates. The uncertainties related to this choice will be discussed in detail
elsewhere, but they do not affect the conclusions drawn in the present study. 

\begin{figure}[ht]
 \includegraphics[width=0.49\textwidth]{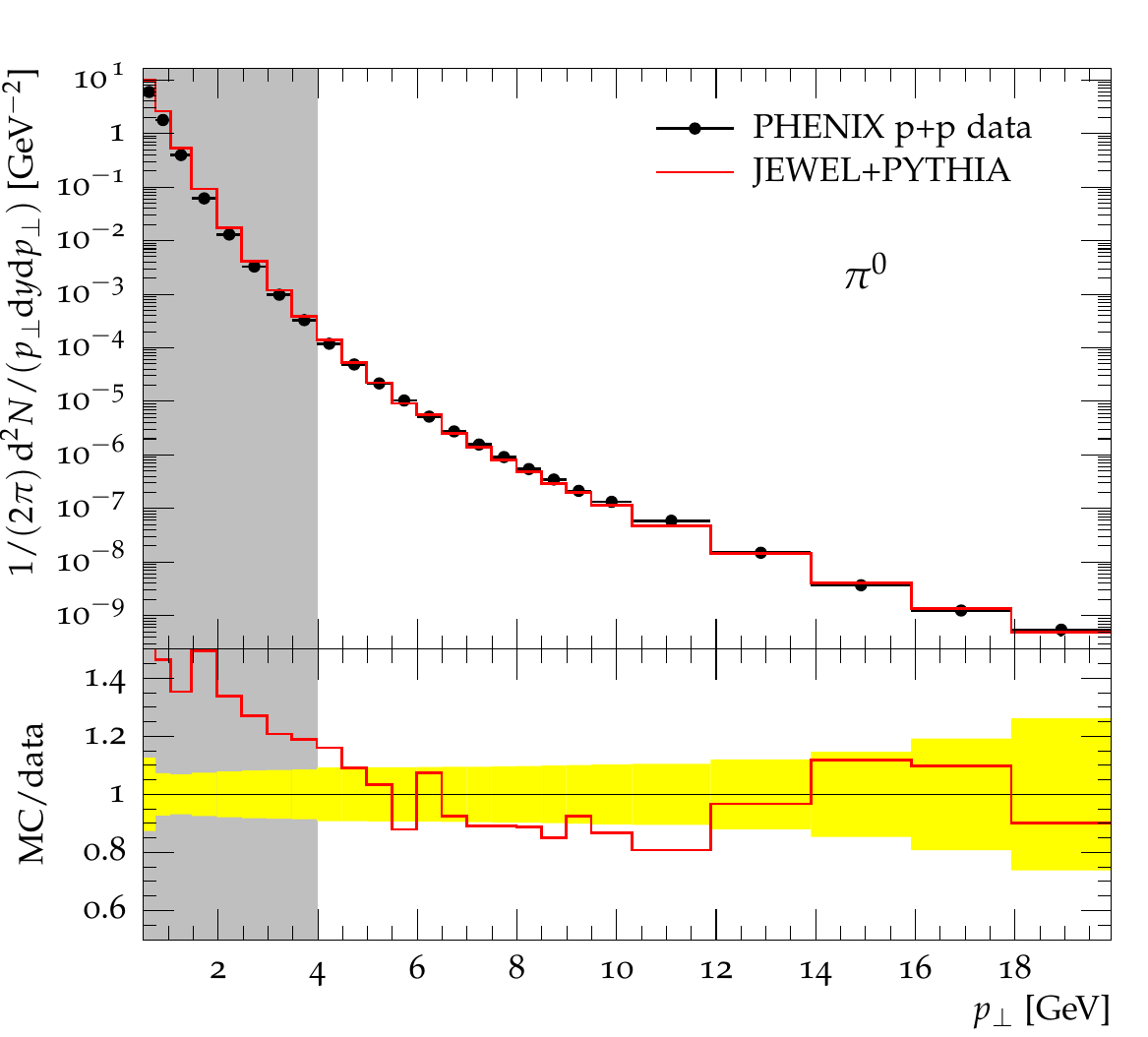}
 \caption{Neutral pion spectrum in $pp$ collisions at $\sqrt{s}=\unit[200]{GeV}$ simulated with \textsc{Jewel+Pythia}
and compared to \textsc{Phenix} data~\cite{Adare:2007dg}. The Analysis of Monte Carlo events and plots were made with
Rivet~\cite{Buckley:2010ar}.}
 \label{fig::pi0spec}
\end{figure}

\medskip
In the absence of a medium the \jewel parton shower reduces to a standard (vacuum) parton
shower. 
The parton shower implementation in \jewel was thoroughly tested in $e^+e^-$ collisions at 
\textsc{Lep} and $pp$ reactions at \textsc{Rhic} and \textsc{Lhc}. As an example for the
generally good performance, Fig.~\ref{fig::pi0spec} shows a comparison with the 
neutral pion spectrum measured by \textsc{Phenix}~\cite{Adare:2007dg}, which forms the
baseline for the measurement of the nuclear modification factor $R_{\text{AA}}$.
Above $\pt \simeq \unit[4]{GeV}$, the \textsc{Jewel+Pythia} results agree with the data 
on a level of roughly \unit[15]{\%} over about 6 decades.

\begin{figure}[ht]
 \includegraphics[angle=-90,width=0.48\textwidth]{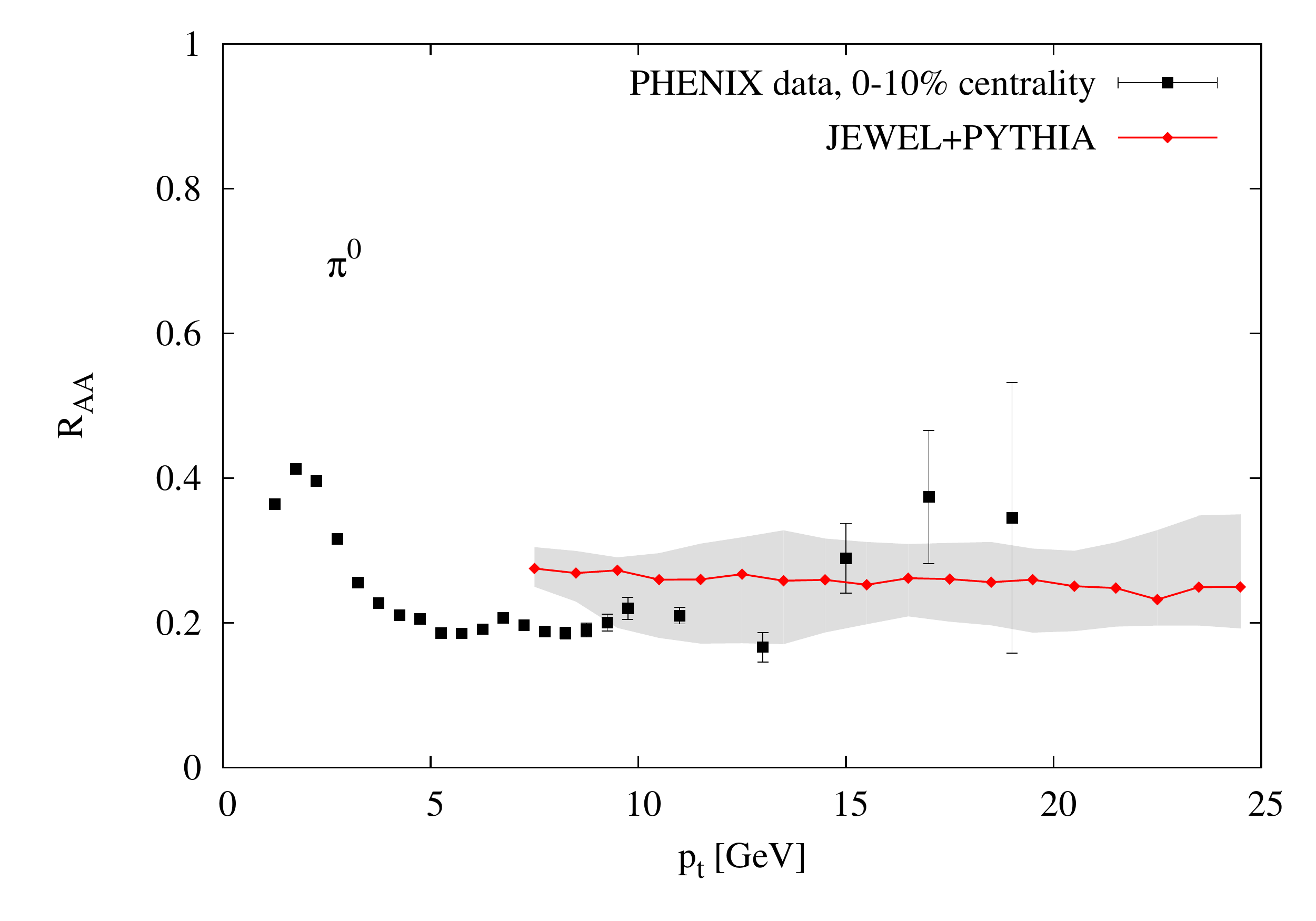}
 \caption{Nuclear modification factor for neutral pions in Au+Au collisions at $\sqrt{s}=\unit[200]{A\,GeV}$ in the
\unit[0-10]{\%} centrality class simulated with \textsc{Jewel+Pythia} and compared to \textsc{Phenix}
data~\cite{Adare:2008qa} (only statistical errors shown). The grey band indicates a variation of the
Debye mass by $\pm\unit[10]{\%}$.}
 \label{fig::phenixraa}
\end{figure}

Turning to medium-modifications of hadron spectra, we fix the critical temperature $\tc = \unit[165]{MeV}$, 
consistent with the expected temperature at which QCD matter transfers from partonic to hadronic degrees
of freedom. To obtain a fair agreement with the measured nuclear modification factor at \textsc{Rhic}, 
\textsc{Jewel+Pythia} requires an initial temperature of $\ti=\unit[350]{MeV}$ at initial proper time 
$\taui=\unit[0.8]{fm}$, see Fig.~\ref{fig::phenixraa}. This is remarkably consistent with the input 
parameters in fluid dynamic simulations of heavy ion collisions. We note that at high $\pt$, where the 
Monte Carlo results are reliable, they reproduce both the factor $\sim 5$ suppression and the approximately flat 
$\pt$-dependence seen in the data. Varying the Debye mass by $\pm \unit[10]{\%}$ has a significant effect on the 
overall suppression, indicated by the grey band, but hardly affects the shape. 
Other sources of uncertainties like pdf-uncertainties, choice of initial time
$\taui$ etc. can be expected to be smaller and will be discussed elsewhere.

\begin{figure}[ht]
 \includegraphics[angle=-90,width=0.48\textwidth]{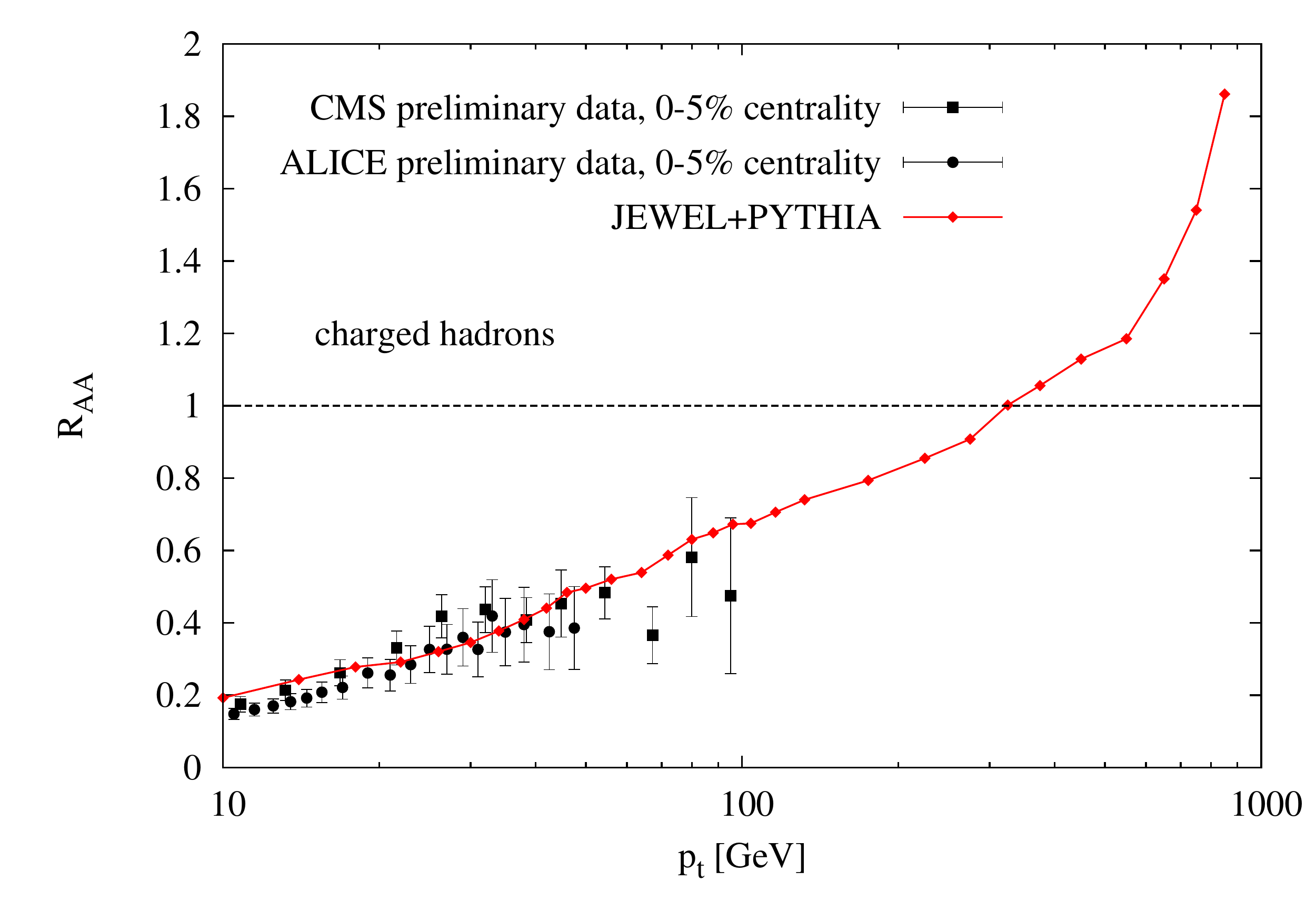}
 \caption{Nuclear modification factor for charged hadrons in Pb+Pb collisions at $\sqrt{s}=\unit[2.76]{A\,TeV}$ in the
\unit[0-5]{\%} centrality class simulated with \textsc{Jewel+Pythia} and compared to preliminary
\textsc{Cms}~\cite{Lee:2011cs} and \textsc{Alice}~\cite{Appelshauser:2011ds} data (only maximum of
statistical and systematic and errors shown).}
 \label{fig::cmsraa}
\end{figure}

The charged hadron multiplicities measured in heavy ion collisions constrain the initial entropy 
density of the system, 
$s_{\text{i}} \taui \propto \d N/\d y$, $s_{\text{i}} \propto \epsilon_{\text{i}}/\ti \propto \ti^3$
and therefore allow to relate the initial temperatures at \textsc{Rhic} and at the \textsc{Lhc},
\begin{equation}
 \ti^{\text{LHC}} = \ti^{\text{RHIC}} \left( \frac{\taui^{\text{RHIC}}}{\taui^{\text{LHC}}} \frac{\left.
  {\d N / \d y} \right|_{\text{LHC}}}{\left. {\d N /\d y}\right|_{\text{RHIC}}} \right)^{1/3} \,. 
\end{equation}
The observation of a factor 2.2 increase in the event multiplicity from \textsc{Rhic} to \textsc{Lhc} is therefore
consistent with an initial temperature $\ti=\unit[530]{MeV}$ at $\taui=\unit[0.5]{fm}$ at the \textsc{Lhc}. 
There is some freedom in initializing the fluid dynamic evolution at the \textsc{Lhc} at a different
initial time $\taui$, but this is numerically unimportant. At early times the parton shower is
dominated by emissions at rather high scales initiated by the initial hard scattering, and this 
high virtuality protects the partons from medium-induced emissions and makes them insensitive to 
the medium at early times. Thus, the medium at the \textsc{Lhc} is specified in terms of parameters
fixed in Fig.~\ref{fig::phenixraa}. As seen from Fig.~\ref{fig::cmsraa}, the calculation
of \textsc{Jewel+Pythia} then leads without any additional adjustments to a very good agreement 
with preliminary data of the nuclear modification factor at the \textsc{Lhc}.

To understand the characteristically different $\pt$-dependencies of $R_{\text{AA}}$ at \textsc{Rhic}
and at the \textsc{Lhc}, we have performed control simulations in which a \textsc{Lhc}-like
distribution of hard processes is fragmented in a \textsc{Rhic}-like medium, and vice versa (data not
shown). This showed that the significant change in the slope of $R_{\text{AA}}(\pt)$ from \textsc{Rhic} to
the \textsc{Lhc} can be attributed fully to the $\sqrt{s}$-dependence of the distribution of initial hard processes.
We therefore conclude that a purely perturbative dynamics of parton energy loss supplemented by an
arguably simple model of the medium whose characterisation matches physical expectations, 
can account for the main features of the measured nuclear modification factors, including the
strength of the suppression pattern, and its $\sqrt{s}$- and $\pt$-dependence. 

At very large $\pt$ the nuclear modification factor continues 
to rise above unity. This is a purely 
kinematical effect that becomes visible at very large $\pt$ where the energy loss
starts to vanish. The elastic scattering of energetic partons converts 
longitudinal momentum into transverse momentum and thus  effectively makes the 
$\pt$-spectrum harder. While this is a generic effect its size and the turn-on 
point to some degree dependend on the medium model, as 
they are sensitive to the amount of scattering centres encountered in the 
forward direction.

\medskip

In this publication we have presented a novel description of jet quenching dynamics entirely 
based on standard perturbative technology also used in the simulation of proton--proton 
collisions, and 
implemented it into the Monte Carlo generator \textsc{Jewel}. In
contrast to other similar attempts, we do not only overcome certain limitations imposed by
the eikonal limit
\footnote{Although the present code goes far beyond the eikonal kinematics, it is clear how to recover this limit of the full emission pattern encoded here.}, 
but we also resolve the dichotomy in the description of in-medium and 
vacuum radiation, basing all emissions on the same parton shower. We reinterpret
induced radiation as regular, although infra-red-regulated, $2\to 2$ parton scatterings,
supplemented with our parton shower. This of course makes any distinction of elastic and 
inelastic scattering obsolete. We hinted at various ways of how our model can systematically
be improved, using a well--\-understood perturbative language. This framework is flexible
enough to accommodate observables such as correlations, or jets, which reach far beyond 
the very inclusive ones studied here. For the fairly successful simulation of the nuclear 
modification factor at \textsc{Rhic} and \textsc{Lhc} we used a simple Bjorken model. 
Taking the description of this inclusive data as a proof of principle, we reserve further 
and more detailed investigations to future work.

\bibliographystyle{apsrev4-1.bst}
\bibliography{jetquenching}

\end{document}